\documentclass[preprint,amsmath,amssymb,amsfonts,showpacs]{revtex4}
\usepackage{txfonts}

\def\ra{\rangle}
\def\la{\langle}
\def\bb{\mathbb}

\def\be{\begin{equation}}
\def\ee{\end{equation}}
\def\ba{\begin{array}}
\def\ea{\end{array}}

\def\qed{\leavevmode\unskip\penalty9999 \hbox{}\nobreak\hfill
     \quad\hbox{\leavevmode  \hbox to.77778em{%
               \hfil\vrule   \vbox to.675em%
               {\hrule width.6em\vfil\hrule}\vrule\hfil}}
     \par\vskip3pt}
\newtheorem{theorem}{Theorem}

\newtheorem{cor}[theorem]{Corollary}

\begin{document}

\begin{center} \bf\large
Faithful Teleportation with Arbitrary Pure or Mixed Resource States
\end{center}
\vskip 1mm

\begin{center}
Ming-Jing Zhao$^{1}$, {Zong-Guo Li$^{2}$}, {Shao-Ming Fei$^{1,3}$},
{Zhi-Xi Wang$^{1}$}, {Xianqing Li-Jost$^{3}$} \vspace{2ex}

\begin{minipage}{5.3in}

\small $~^{1}$ {School of Mathematical Sciences, Capital Normal
University, Beijing 100048, China}

{\small $~^{2}$ College of Science, Tianjin University of
Technology, Tianjin, 300191, China}

{\small $~^{3}$ Max-Planck-Institute for Mathematics in the
Sciences, 04103, Leipzig, Germany}

\end{minipage}
\end{center}

\vskip 4mm

{\centerline{\large Abstract}}

\begin{center}
\begin{minipage}{5.6in}

We study faithful teleportation systematically with arbitrary entangled states as resources. The necessary
conditions of mixed states to complete perfect teleportation are
proved. Based on these results, 
the necessary and sufficient conditions of
faithful teleportation of an unknown state $|\phi\rangle$ in ${\bb
C}^d$ with entangled source $\rho$ in ${\bb C}^m \otimes {\bb
C}^d $ and ${\bb C}^d \otimes {\bb C}^n $ are derived. It is shown
that for $\rho$ in ${\bb C}^m \otimes {\bb C}^d $, $\rho$ must be a
maximally entangled state, while for $\rho$ in $ {\bb C}^d
\otimes{\bb C}^n $, $\rho$ must be a pure maximally entangled state.
Moreover, we show that the sender's measurements must be all projectors of
maximally entangled pure states. The relations
between the entanglement of formation of the resource states and faithful teleportation
are also discussed.
\bigskip
PAC numbers: {03.67.Mn, 03.65.Ud}

\end{minipage}
\end{center}

\section{Introduction}

Quantum teleportation plays an important role in quantum information
processing. It,
employing classical communication and shared resource of
entanglement, allows to transmit an unknown quantum state from a
sender to a receiver that are spatially separated. Bennett {\it et.
al.} \cite{first tele} first demonstrated the teleportation of an arbitrary
qubit state in terms of an entangled
Einstein-Podolsky-Rosen pair. Generally it has proved that only maximally entangled pure states in
${\bb C}^d\otimes {\bb C}^d$ could faithfully teleport an arbitrary pure state
in ${\bb C}^d$ \cite{fefandtel, fef tel alb}.

Multipartite states have been also used in
faithful and deterministic teleportation.
For instance, the three-qubit GHZ state and a class of
$W$ states can be used for faithful teleportation of one qubit state
\cite{ghz tele, w tele}. In fact, they are all maximally entangled pure states in ${\bb C}^2
\otimes {\bb C}^2$ between Alice and Bob. Some five-qubit state is
capable of faithful teleportation of arbitrary two-qubit states
\cite{five-qubit tele}. The tensor
products of two Bell states \cite{G. Rigolin} and the genuine
four-qubit entangled states \cite{Y. Yeo} are also used for faithful
teleportation of two-qubit states, in which the teleportation channels can
be regarded as maximally entangled states in
${\bb C}^4 \otimes {\bb C}^4$ shared by the sender and receiver.
Basically multipartite pure entangled states are analogous
to higher dimensional bipartite pure states in quantum teleportation.

In a realistic case, due to the decoherence the pure
maximally entangled states may evolve into mixed entangled
states, which could make the teleportation of a state in ${\bb C}^d$ imperfect, if the shared
entangled state is in ${\bb C}^d \otimes {\bb C}^d $.

In this paper, we study faithful teleportation with high dimensional bipartite pure or mixed
states as the resource.
We investigate faithful teleportation of an arbitrary pure state
$|\phi\ra$ in ${\bb C}^d$ by using entangled state
$\rho$ in ${\bb C}^m \otimes {\bb C}^n$ $(m, n \geq d)$.
We first derive the necessary conditions of mixed states as entangled resources
to fulfill faithful teleportation. The necessary and sufficient conditions for
faithful teleportation are obtained for resource states $\rho$ in ${\bb C}^m \otimes {\bb
C}^d $ and ${\bb C}^d \otimes {\bb C}^n $. We show that
$\rho$ in ${\bb C}^m \otimes {\bb C}^d $, either pure or mixed,
must be maximally entangled. While $\rho$ in ${\bb C}^d \otimes {\bb C}^n $ must be
pure maximally entangled. Moreover, to fulfill faithful teleportation, the sender's
measurements must be all projectors of maximally entangled pure
states.  For $\rho$ in ${\bb C}^m \otimes {\bb C}^n $, $m,\,n>d$, we present
some classes of states for faithful teleportation.

\section{Teleportation with entangled resource states in ${\bb C}^m \otimes {\bb C}^n $}

Let $|\phi\ra= \sum_{i=1}^d \alpha_i|i\ra$ be the unknown pure state that is to be
sent from Alice to Bob, where $\{|i\rangle\}_{i=1}^d$ is the orthonormal basis of ${\bb C}^d $.
Let $\rho$ be the entangled state shared by Alice and Bob. To carry out teleportation Alice needs
to perform projective measurements on her two particles: one in state
$|\phi\rangle$ and one part of the entangled state $\rho$. Learning
the measurement results from Alice via the classical communication
channel, Bob applies a corresponding unitary transformation on
the other part of the entangled state $\rho$, so as to transform
the state of this part to the unknown state
$|\phi\rangle$. In the following we investigate this traditional faithful teleportation
with arbitrary dimensional bipartite entangled state $\rho$.

We first consider the case that $\rho$ in ${\bb C}^m \otimes {\bb C}^n$ $(m, n \geq d)$.

\begin{theorem}\label{th1}
If a mixed state $\rho=\sum_i p_i |\psi_i\rangle \langle \psi_i |$, $p_i\geq0$, $\sum_{i=1}^k p_i=1$,
is an ideal resource for faithful teleportation, then every pure state $|\psi_i\rangle$ is
the ideal resource for faithful teleportation.
\end{theorem}

Proof. Let $\{| \tilde\psi_j\ra \}_{j=1}^{r}$ be an orthonormal basis in ${\bb
C}^r$, $\sum_{j=1}^{r} |\tilde\psi_j\ra \la \tilde\psi_j|=I_{r}$,
where $I_{r}$ is the $r \times r$ identity matrix.
Assume Alice makes a complete projective measurement $\{| \tilde\psi_j\rangle \langle
\tilde\psi_j| \}_{j=1}^{r}$ on the initial state $|\phi\ra \la\phi|\otimes \rho$,
where $|\phi\rangle$ is unknown state to be teleported. Then Bob applies
the unitary operation $U^{(j)\dagger}$ on his part with respect to
Alice's measurement result $j$. A faithful teleportation implies that
\be\label{combined state}
|\phi\ra \la \phi| \otimes \rho=\sum_{i=1}^k p_i |\phi\rangle
\langle \phi| \otimes |\psi_i\rangle \langle\psi_i|
=\sum_{j,j^\prime=1}^{r} q_j |\tilde\psi_j\ra \la
\tilde\psi_{j^\prime}| \otimes U^{(j)}|\phi\ra \la \phi|
U^{(j^\prime)\dagger}
\ee
with $\sum_{j=1}^{r} q_j =1$, $q_j \geq 0$, $j=1, \cdots,r$. After
Alice's complete projective measurement, from Eq. (\ref{combined state}) we obtain
\be\label{to use rank}
\sum_{i=1}^k p_i \la \tilde\psi_{j}|(|\phi\ra  \la\phi | \otimes
|\psi_i\ra \la\psi_i|) |\tilde\psi_j\ra
= q_j U^{(j)}|\phi\ra \la \phi| U^{(j)\dagger}
\ee
and
\be\label{u^j invariant}
\la \tilde\psi_{j}| (|\phi\ra \la\phi| \otimes |\psi_i\ra
\la\psi_i|)| \tilde\psi_j\ra= f_{ij} U^{(j)}|\phi\ra \la \phi|
U^{(j)\dagger},
\ee
with $\sum_{i=1}^k \sum_{j=1}^{r} f_{ij}=1$, $f_{ij}\geq 0$, $i=1,
\cdots, k$, $j=1, \cdots, r$. From Eq. (\ref{u^j invariant}) we have
\begin{eqnarray}\label{every pure tele}
\sum_{j=1}^{r} \la \tilde\psi_{j}|( |\phi\ra \la\phi |\otimes
|\psi_i\ra \la\psi_i|)
 |\tilde\psi_j\ra= \sum_{j=1}^{r} f_{ij} U^{(j)}|\phi\ra \la
\phi| U^{(j)\dagger}
\end{eqnarray}
for $i=1, \cdots, k$. Due to the completeness of the projectors $\{
|\tilde\psi_j\rangle \langle  \tilde\psi_j|\}$, $f_{ij}$ satisfies
$\sum_{j=1}^r f_{ij}=1$ for each $i$. Therefore
every pure state $|\psi_i\ra$ must be an ideal resource for faithful
teleportation. \qed

From the theorem one has the following
necessary condition of mixed states for faithful teleportation.
\begin{cor}\label{cor}
If mixed state $\rho$ is the ideal resource for faithful
teleportation, then its eigenstates must be all ideal resource for
faithful teleportation.
\end{cor}

Utilizing the necessary conditions of mixed states, we consider now
which kind of states in ${\bb C}^m \otimes {\bb C}^n$ can be used
as entangled resource for faithful teleportation of an unknown state
$|\phi\rangle$ in ${\bb C}^d$. We systematically study the problem in four cases.

{\em Case} \textrm{i}). {\em $n=d$, pure states}

Since any pure state $|\psi\rangle$ in ${\bb C}^m \otimes {\bb C}^d$ can be transformed
into some pure state in ${\bb C}^d \otimes {\bb C}^d$ under local
unitary transformations, we only need to consider pure states $|\psi\rangle$ in ${\bb C}^d \otimes {\bb
C}^d$. In this case, it has been shown that only maximally entangled pure
states can be used for faithful teleportation in ${\bb C}^d \otimes
{\bb C}^d$, if the Bell measurements are applied by Alice \cite{first tele, fefandtel, fef tel alb}.
Here we give an alternative proof of this result for consistency and the use for the rest of this paper.
In addition, from the proof we show that, to fulfill faithful teleportation, Alice's measurements must be
projectors of maximally entangled pure states.

\begin{theorem}\label{th2}
A pure state $|\psi\rangle$ in ${\bb C}^d \otimes {\bb C}^d$ is an
ideal resource for faithful teleportation of $|\phi\ra= \sum_{i=1}^d
\alpha_i|i\ra$, if and only if $|\psi\rangle$ is maximally entangled.
Moreover, Alice's measurements must be all projectors of
maximally entangled states.
\end{theorem}

Proof. Let $|\psi\ra=\sum_{i,j=1}^d a_{ij} |ij\ra$ be the entangled
pure state shared by Alice and Bob. To teleport the unknown state
$|\phi\ra$, Alice carries out complete measurements
$\{|\psi_{st}\rangle \langle \psi_{st}|\}_{s,t=1}^{d}$, where
$|\psi_{st}\ra=\sum_{p,q=1}^d U_{pq,st}|pq\ra$ satisfying $\langle
\psi_{s^\prime, t^\prime} |\psi_{s,t}\rangle= \sum_{p,q=1}^d
U_{pq,st}U_{pq,s^\prime t^\prime}^*=\delta_{s,s^\prime}
\delta_{t,t^\prime} $. One has
\begin{eqnarray}
\label{ddpure}
|\phi\ra |\psi\ra
&=& \sum_{i,j,k=1}^d \alpha_i a_{jk} |ijk\ra
=\sum_{s,t=1}^d \sum_{i,j,k=1}^d \alpha_i a_{jk} |\psi_{st}\ra \la \psi_{st}|ijk\ra \nonumber\\
&=& \sum_{s,t=1}^d |\psi_{st}\ra (\sum_{i,j,k=1}^d U_{ij,st}^*
\alpha_i a_{jk} |k\ra )
=\sum_{s,t=1}^d |\psi_{st}\ra
A^T V_{st}^\dagger |\phi\ra,
\end{eqnarray}
where $(V_{st})_{ij}=(U_{ij,st})$ and $(A)_{ij}=(a_{ij})$ are the coefficient
matrices of $|\psi_{st}\ra $ and $|\psi\ra$ respectively. If
$|\psi\rangle$ is an ideal resource for faithful teleportation, then
$A^T V_{st}^\dagger$ should be unitary up to a constant factor:
\begin{eqnarray}\label{A V unitary}
A^T V^\dagger_{st} = c_{st} W_{st},
\end{eqnarray}
for some unitary matrix $W_{st}$ and $0\leq c_{st}\leq1$, $s, t =1, \cdots, d$,
\begin{eqnarray}\label{probility}
\sum_{s,t=1}^d |c_{st}|^2=1.
\end{eqnarray}
Let $A^T=U_1 \Lambda_1 V_1$ and $V_{st}^\dagger=U_{2,st}
\Lambda_{2,st} V_{2,st}$ be the singular decompositions of $A^T$ and
$V_{st}^\dagger$ respectively, where $U_1, V_1,U_{2,st},
V_{2,st}$ are unitary matrices, $\Lambda_1=diag(\lambda_1,\cdots,\lambda_d)$,
$\Lambda_{2,st}=diag(\mu_{1,st},\cdots,\mu_{d,st})$, $\lambda_i$ and $\mu_{i,st}$ are
the corresponding singular values. Due to the
normality, $tr AA^\dagger=tr V_{st} V_{st}^\dagger =1$, one gets
\begin{eqnarray}
\sum_{i=1}^d |\lambda_i|^2 &=& 1, \label{sum lambda 1}\\
\sum_{i=1}^d |\mu_{i,st}|^2&=&1 \label{sum mu 1}.
\end{eqnarray}
Then from Eq. (\ref{A V unitary}) we have
\begin{eqnarray}
c_{st}^2I_d= A^T V_{st}^\dagger V_{st} A^{T\dagger}
=U_1 \Lambda_1 V_1U_{2,st} \Lambda_{2,st} \Lambda_{2,st}^\dagger
U_{2,st}^\dagger V_1^\dagger \Lambda_1^\dagger U_1^\dagger
\end{eqnarray}
and
\begin{eqnarray}
c_{st}^2 (\Lambda_1 \Lambda_1^\dagger)^{-1} = V_1U_{2,st}
\Lambda_{2,st} \Lambda_{2,st}^\dagger U_{2,st}^\dagger V_1^\dagger,
\end{eqnarray}
which give rise to
\begin{eqnarray}\label{lambda i and mu i 1}
|\lambda_i|^{-2} = \frac{1}{c_{st}^2} |\mu_{i,st}|^2
\end{eqnarray}
and
\begin{eqnarray}\label{lambda i and mu i 2}
|\lambda_i|^2 = \frac{c_{st}^2}{ |\mu_{i,st}|^2}
\end{eqnarray}
by reordering $\{ \lambda_i \}$ and $\{ \mu_{i,st} \}$.
From Eq. (\ref{sum lambda 1}) and Eq. (\ref{lambda i and mu i 1}) we have
$$
\frac{1}{c^2_{st}} \sum_{i=1}^d |\mu_{i,st}|^2 = \sum_{i=1}^d |\lambda_i|^{-2} \leq d^2 .
$$
Using Eq. (\ref{sum mu 1}) one has
$ c^2_{st} \geq 1/d^2$
for $s,t=1, \cdots, d$, which results in
\begin{eqnarray}\label{condition c}
c^2_{st} = \frac{1}{d^2}
\end{eqnarray}
by taking into account Eq. (\ref{probility}).
Inserting Eq. (\ref{condition c}) into Eq. (\ref{lambda i and mu i 2}) and using Eq. (\ref{sum lambda 1}) we get
\begin{eqnarray}\label{mu i d^2}
\sum_{i=1}^d \frac{1}{|\mu_{i,st}|^2} =d^2.
\end{eqnarray}
Hence in terms of Eq. (\ref{sum mu 1}), Eq. (\ref{lambda i
and mu i 2}) and Eq. (\ref{mu i d^2}) we obtain
\begin{eqnarray}\label{equal shcmidt coef}
|\lambda_i|^2 =|\mu_{i,st}|^2= \frac{1}{d},~~~i,s,t=1,\cdots, d,
\end{eqnarray}
which are just the square of the Schmidt
coefficients of $|\psi\ra$ and $|\psi_{st}\rangle$ respectively.
Therefore $A^T=\frac{1}{\sqrt{d}}\tilde U$ and
$V_{st}=\frac{1}{\sqrt{d}}\tilde V_{st}$ for some unitary matrices
$\tilde U$ and $\tilde V_{st}$. As a result, the
shared entangled state $|\psi\ra$ and the state $|\psi_{st}\rangle$ in the projective measurements
$\{|\psi_{st}\rangle \langle \psi_{st}|\}$ must be all maximally entangled ones.
At last, the initial state can be expressed as
\begin{eqnarray}
|\phi\ra |\psi\ra =\frac{1}{d}\sum_{s,t=1}^d |\psi_{st}\ra \tilde U_{st} |\phi\ra,
\end{eqnarray}
where $\tilde U_{st}=d A^T V_{st}^\dagger$ is determined by the
shared state $|\psi\ra$ and the projective measurement operators
$|\psi_{st}\rangle \langle \psi_{st}|$. To carry out the teleportation, Alice
measures her particles by $d^2$ orthonormal projectors, and informs
Bob the measurement results. Each result appears in her measurements
with probability $\frac{1}{d^2}$. According to Alice's measurement
result $st$, Bob fulfills faithful teleportation by applying unitary
operation $\tilde U_{st}^\dagger$ on his part of the entangled
resource.

Conversely, if the shared entangled state is maximally
entangled, it is straightforward to prove that faithful teleportation can be carried out.
A maximally entangled state can be generally expressed as
$U_1\otimes U_2|\psi\rangle$, where
$|\psi\rangle=\frac{1}{\sqrt{d}}\sum_{i=1}^d|ii\rangle$, $U_1$ and $U_2$ are unitary matrices.
Taking $|\psi\rangle$ as the entangled resource to teleport
$|\phi\rangle=\sum_{i=1}^d\alpha_i|i\rangle$, one has
\begin{eqnarray*}
|\phi\rangle \otimes |\psi\rangle =
\frac{1}{d}\sum_{s,t=1}^d|\psi_{st}\rangle\otimes
U_{st}|\phi\rangle,
\end{eqnarray*}
where $|\psi_{st}\rangle=\frac{1}{\sqrt{d}}U_{st}^\dagger\otimes
I|\psi\rangle$, $s, t=1,2, \cdots, d$. Here $\{U_{st}\}$ is the
basis of the unitary operators satisfying $tr(U_{st}U_{s^\prime
t^\prime}^\dagger)=d \delta_{ss^\prime} \delta_{tt^\prime}$ and
$tr(U_{st}U_{s t}^\dagger)=I_d$. For instance, one could choose
$U_{st}=h^tg^s$ with $d\times d$ matrices $h$ and $g$ such that
$h|j\rangle = |(j + 1)\mod d\rangle$, $g|j\rangle = \omega ^j
|j\rangle$, $\omega = \exp\{-2i\pi/d\}$, $s,t=1,2,\cdots, d$, as the
basis of the unitary operators to perform the faithful
teleportation. Faithful teleportation with other maximally entangled
pure states can be similarly analyzed. \qed

{\em Case} \textrm{ii}). {\em $n=d$, mixed states}

\begin{theorem}\label{th3}
A mixed state $\rho$ in ${\bb C}^m \otimes {\bb C}^d$ with rank $k$
can be used as the entangled resource for faithful teleportation of $|\phi\ra= \sum_{i=1}^d
\alpha_i|i\ra$, if and only if $\rho$ is the mixed maximally
entangled state \cite{lizhao}: $\rho= \sum_{x=1}^k p_x |\psi_x\rangle \langle
\psi_x|$, where $|\psi_x\rangle$ is maximally entangled in $H_x
\otimes {\bb C}^d$, $\dim H_x =d$, $x=1,\cdots, k$, $\{H_x\}$ are complex vector spaces that are
orthogonal to each other.
\end{theorem}

One of the maximally entangled state in ${\bb C}^m \otimes {\bb C}^d$ ($m\geq d$) is of the form,
$|\phi\rangle\!\!=\!\!\sum_{i=1}^d\frac{1}{\sqrt{d}}|ii\rangle$. Quantified by a certain
entanglement measure, a mixed maximally entangled state
has the same degree of entanglement as this pure maximally entangled state.
This fact holds true for any entanglement measures that does not increase under local
operations and classical communications in literature \cite{lizhao}.

Proof. It has been proved that the mixed maximally entangled state
\cite{lizhao} can be used to fulfill faithful teleportation. Now we prove
the converse. Suppose the mixed state $\rho$ with rank $k$ in ${\bb
C}^m \otimes {\bb C}^d$ is the ideal resource for teleportation. From corollary \ref{cor} we
have that the $k$ orthogonal eigenstates $\{|\psi_i\rangle\}$ of $\rho$ with respect to
nonzero eigenvalues are all maximally entangled. In fact, they
can be constructed in the following way. We assume, without loss of
generality, $|\psi_1\ra$ be maximally entangled in $H_1 \otimes {\bb
C}^d$ with $H_1={\bb C}^d= S\!\!pan\{|1\rangle, \cdots,
|d\rangle\}$, where $\{|i\rangle\}_{i=1}^d$ is an orthonormal basis of ${\bb C}^d $.
From Eq. (\ref{every pure tele}) one has
\begin{eqnarray}
\sum_{j=1}^{d^2} \la \tilde\psi_{j}| (|\phi\ra \la\phi |\otimes
|\psi_1\ra \la\psi_1|) |\tilde\psi_j\ra= \sum_{j=1}^{d^2} f_{1j}
U^{(j)}|\phi\ra \la \phi| U^{(j)\dagger},
\end{eqnarray}
where $f_{1j}={1}/{d^2}$ for $j=1, \cdots, d^2$, and $f_{1j}=0$
for $j>d^2$. Here $\{ |\tilde\psi_j\ra \}_{j=1}^{d^2}$ are maximally
entangled states constituting an orthonormal basis in ${\bb
C}^d\otimes H_1$. Similarly state  $|\psi_2\ra$ satisfies
\begin{eqnarray}
\sum_{j=d^2+1}^{2d^2} \la \tilde\psi_{j}| (|\phi\ra \la\phi|\otimes
|\psi_2\ra \la\psi_2|)  |\tilde\psi_j\ra= \sum_{j=d^2+1}^{2d^2}
f_{2j} U^{(j)}|\phi\ra \la \phi| U^{(j)\dagger}
\end{eqnarray}
with $f_{2j}= {1}/{d^2}$ for $j=d^2+1, \cdots, 2d^2$, and $f_{2j}=0$
otherwise. $\{ |\tilde \psi_j\ra \}_{j=d^2+1}^{2d^2}$ are maximally
entangled and constitute an orthonormal basis in ${\bb C}^d \otimes
H_2$, $\dim H_2=d$. Hence
$|\psi_2\ra$ is maximally entangled in $H_2\otimes {\bb C}^d$.
From the orthogonality of $\{ |\tilde \psi_j\ra
\}_{j=1}^{d^2}$ and $\{ |\tilde \psi_j\ra \}_{j=d^2+1}^{2d^2}$, we
know that $H_2$ is orthogonal to $H_1$. Other eigenstates $|\psi_x\ra$, $2\leq x\leq k$, can be treated
similarly,
\begin{eqnarray}\label{every dd pure tele}
\!\!\!\!\sum_{j=(x-1)d^2+1}^{xd^2}\!\!\!\! \la \tilde\psi_{j}|
(|\phi\ra \la\phi |\otimes |\psi_x\ra \la\psi_x| )|\tilde\psi_j\ra=
\!\!\!\!\sum_{j=(x-1)d^2+1}^{xd^2}\!\!\!\! f_{xj} U^{(j)}|\phi\ra
\la \phi| U^{(j)\dagger}.
\end{eqnarray}
$\{ |\tilde \psi_j\ra \}_{j=(x-1)d^2+1}^{xd^2}$ are maximally
entangled and constitute the orthonormal basis in ${\bb C}^d\otimes
H_x$, where $\{H_x\}$ are orthogonal to each other, $\dim H_x=d$ for
$x=1, \cdots, k$. Hence $|\psi_x\ra$ is maximally entangled in
$H_x\otimes {\bb C}^d$. From the above analysis, we have $m\geq kd$
with $k$ the rank of $\rho$. The probability of the outcomes of each measurement
depends on the eigenvalues $p_i$, $i=1, \cdots, k$. Therefore
a mixed state $\rho$ in ${\bb C}^m\otimes {\bb C}^d$ that can be
used for faithful teleportation of $|\phi\ra$ in ${\bb C}^d$ must be
a mixed maximally entangled state. \qed

As an example, we consider the faithful teleportation of
$|\phi\rangle=\alpha|0\rangle +\beta|1\rangle$ by using a mixed
maximally entangled state $\rho_0$ in ${\bb C}^4\otimes {\bb C}^2$,
$\rho_0=\frac{1}{2}|\psi_1^+\rangle
\langle \psi_1^+| + \frac{1}{2}|\psi_2^+\rangle \langle \psi_2^+|$
with $|\psi_1^+\rangle=\frac{1}{\sqrt{2}}(|00\rangle +|11\rangle)$
and $|\psi_2^+\rangle=\frac{1}{\sqrt{2}}(|20\rangle +|31\rangle)$.
By straightforward calculations one has
\begin{eqnarray*}
|\phi\rangle\langle\phi| \otimes \rho_0
=&&\frac{1}{8}(|\psi_1^+\rangle \otimes|\phi\rangle +
|\psi_1^-\rangle \otimes \sigma_3|\phi\rangle + |\phi_1^+\rangle
\otimes \sigma_1 |\phi\rangle + |\phi_1^-\rangle \otimes
\sigma_2|\phi\rangle)\\
&&(\langle\psi_1^+| \otimes\langle\phi| + \langle\psi_1^-| \otimes
\sigma_3\langle\phi| + \langle\phi_1^+| \otimes \sigma_1
\langle\phi|-\langle\phi_1^-| \otimes \sigma_2\langle\phi|) \\
&&+ \frac{1}{8}(|\psi_2^+\rangle \otimes|\phi\rangle +
|\psi_2^-\rangle \otimes \sigma_3|\phi\rangle + |\phi_2^+\rangle
\otimes \sigma_1 |\phi\rangle + |\phi_2^-\rangle \otimes
\sigma_2|\phi\rangle) \\
&&(\langle\psi_2^+| \otimes\langle\phi| + \langle\psi_2^-| \otimes
\sigma_3\langle\phi| + \langle\phi_2^+| \otimes \sigma_1
\langle\phi| -\langle\phi_2^-| \otimes \sigma_2\langle\phi|),
\end{eqnarray*}
where $|\psi^-_1\rangle =\frac{1}{\sqrt{2}} (|00\rangle -
|11\rangle)$, $|\phi^\pm_1\rangle =\frac{1}{\sqrt{2}} (|01\rangle \pm
|10\rangle)$, $|\psi_2^-\rangle
=\frac{1}{\sqrt{2}} (|20\rangle - |31\rangle)$ and
$|\phi_2^\pm\rangle =\frac{1}{\sqrt{2}} (|21\rangle \pm |30\rangle)$
are maximally entangled states, $\sigma_1=|0\rangle \langle1| + |1\rangle
\langle0|$, $\sigma_2=-|0\rangle \langle1| + |1\rangle \langle0|$ and
$\sigma_3=|0\rangle \langle0| - |1\rangle \langle1|$ are Pauli matrices.
Above relation implies that the faithful teleportation can be carried out
with the mixed maximally entangled state $\rho_0$.

In fact, this mixed state $\rho_0$ could be evolved from a
maximally entangled pure state $|\psi_0\rangle=\frac{1}{2}(|00\rangle
+|11\rangle +|22\rangle +|33\rangle)$. If the second part of the pure state $|\psi_0\rangle$
undergoes a noisy channel with Kraus operators $A_1=
|0\rangle \langle 0| + |1\rangle \langle 1|$ and $A_2= |2\rangle
\langle 2| + |3\rangle \langle 3|$, $|\psi_0\rangle$ becomes $\rho_0=\Lambda\otimes I
(|\psi_0\rangle \langle \psi_0|)$. Usually under noisy channel
a maximally entangled pure state becomes
a mixed one which can be no longer used for faithful teleportation.
Our example shows that this channel does
not influence the capability of the input state for faithful
teleportation of one qubit state.

{\em Case} \textrm{iii}). {\em $m=d$, pure or mixed states}

\begin{theorem}\label{th5}
An entangled state $\rho$ in ${\bb C}^d\otimes {\bb C}^n$ can be used for
faithful teleportation of $|\phi\ra= \sum_{i=1}^d \alpha_i|i\ra$, if
and only if it is a maximally entangled pure state in ${\bb C}^d\otimes {\bb C}^d$.
\end{theorem}

Proof. If $\rho$ with rank $k$ is able to teleport $|\phi\ra$
faithfully, from corollary \ref{cor} its eigenstates should be all ideal resources for
teleportation. Hence according to the proof
of theorem \ref{th3}, $\rho$'s eigenvectors associated with the first subsystem
must be orthogonal to each other. Namely the dimension of the first
subsystem of $\rho$ should be $kd$. Since the dimension of the first subsystem of $\rho$ is $d$, we have $k=1$
and $\rho$ is a maximally entangled pure state. \qed

In the above discussions we have used the traditional teleportation protocol:
Alice makes a projective measurement first, Bob applies 
unitary operations correspondingly then. In fact, if the entangled resource
is in ${\bb C}^d\otimes {\bb C}^n$, Bob performs a
projective measurement first, then more entangled states may be
served as ideal resources for faithful teleportation.

\begin{theorem}
State $\rho$ in ${\bb C}^d\otimes {\bb C}^n$ can be used for
faithful teleportation of state $|\phi\ra= \sum_{i=1}^d \alpha_i|i\ra$, if and only if
$\rho$ is the mixed maximally entangled state: $\rho=
\sum_{x=1}^k p_x |\psi_x\rangle \langle \psi_x|$, where
$|\psi_x\rangle$ is maximally entangled in ${\bb C}^d \otimes H_x$,
$\dim H_x =d$, $x=1,\cdots, k$, $n\geq kd$,
$\{H_x\}$ are complex vector spaces that are orthogonal to each other.
\end{theorem}
Proof. Let Bob perform a complete measurement first.
To carry out faithful teleportation, from theorem \ref{th5} the
post-measurement state in ${\bb C}^d\otimes {\bb C}^{n^\prime}$,
$(n^\prime < n)$, must be a maximally entangled pure state in ${\bb
C}^d \otimes {\bb C}^d$. Since local operations do not increase entanglement,
the initial state $\rho$ with rank $k$ must be a maximally entangled one
in ${\bb C}^d \otimes {\bb C}^d$. This implies that $n\geq kd$ and $\rho=
\sum_{x=1}^k p_x |\psi_x\rangle \langle \psi_x|$ with
$|\psi_x\rangle$ the maximally entangled state in ${\bb C}^d \otimes H_x$,
$\dim H_x =d$, $x=1,\cdots, k$, $\{H_x\}$ are orthogonal to each other.

On the other hand, if Bob's measurement can project
the maximally entangled states in ${\bb C}^d\otimes {\bb C}^n$ to be the above
maximally entangled states in ${\bb C}^d \otimes H_x$, they can be used for faithful
teleportation obviously.
Therefore, state $\rho$ in ${\bb C}^d\otimes {\bb C}^n$ can be used for
faithful teleportation if and only if it is the mixed maximally
entangled and $n\geq kd$. \qed

{\em Case} \textrm{iv}). {\em $m,n>d$, pure or mixed states}

Concerning pure states in $m,n>d$, we introduce a class of states:
\begin{eqnarray}
\label{mn-state}
|\psi\ra =c_1|\psi_1\ra + \cdots + c_{l} |\psi_{l}\ra,
\end{eqnarray}
where $|\psi_p\rangle\in H_p^A\otimes H_p^B$ is a maximally entangled state,
$\{H_p^A\}_{p=1}^l$ are complex vector spaces that are orthogonal to each other, $\dim H_p^A =\dim
H_p^B=n_p \geq d$ for $p=1,\cdots, l$, $\sum_{p=1}^l n_p \leq
\min\{m,~n\}$ and $\sum_{i=1}^l |c_i|^2=1$.
Without loss of generality, we
assume Alice's measurements are given by the projectors $\{|\tilde \psi_{st,p}\ra
\la \tilde \psi_{st,p}|\}$, $s=1,\cdots,d$, $t=1,\cdots,n_p$, $p=1,
\cdots, l$. Here for $p=1, \cdots, l$, $\{|\tilde \psi_{st,p}\ra \la
\tilde \psi_{st,p}|\}$ are projectors onto ${\bb C}^d\otimes
H_p^A$. And $\{|\tilde \psi_{st,p}\ra\}$ are maximally
entangled states constituting an orthonormal basis in ${\bb
C}^d\otimes H_p^A$. We have
\begin{eqnarray}
|\phi\ra|\psi\ra = \frac{1}{\sqrt{dn_p}} \sum_{p=1}^{l} \sum_{s=1}^d
\sum_{t=1}^{n_p} c_p |\tilde \psi_{st,p}\ra \tilde U_{st,p}
|\phi\ra,
\end{eqnarray}
where the unitary matrix $\tilde U_{st,p}$ depends on the shared
resource state $|\psi\ra$ and the measurement operator $|\tilde
\psi_{st,p}\ra \la \tilde \psi_{st,p}|$. The probability of Alice's each measurement outcome is $\frac{|c_p|^2}{dn_p}$.


For example, let us consider the teleportation of a qubit state
$|\phi\rangle=\alpha|0\rangle+\beta|1\rangle$ with nonmaximally
entangled pure state $|\psi\rangle =
\sqrt{a}|\eta\rangle+\sqrt{1-a}|\xi\rangle$, with
$|\eta\rangle=\frac{1}{\sqrt{2}}(|00\rangle +|11\rangle)$ and
$|\xi\rangle = \frac{1}{\sqrt{3}}(|22\rangle
+|33\rangle+|44\rangle)$. We have
\begin{eqnarray*}
|\phi\rangle \otimes |\psi\rangle
&=& \frac{\sqrt{a}}{2}(\frac{1}{\sqrt{2}}(|00\rangle
+|11\rangle) \otimes U_1 |\phi\rangle +
\frac{1}{\sqrt{2}}(|00\rangle-|11\rangle) \otimes U_2 |\phi\rangle \\
&&+\frac{1}{\sqrt{2}}(|01\rangle+|10\rangle) \otimes U_3 |\phi\rangle +
\frac{1}{\sqrt{2}}(|01\rangle-|10\rangle) \otimes U_4 |\phi\rangle  )\\
&&\!+\!\sqrt{\frac{1\!-\!a}{6}}
(\frac{1}{\sqrt{2}}(|02\rangle\!+\!|13\rangle) \!\otimes\! V_1 |\phi\rangle \!+\!
\frac{1}{\sqrt{2}}(|02\rangle\!-\!|13\rangle) \otimes V_2 |\phi\rangle \\
&&+\frac{1}{\sqrt{2}}(|03\rangle+|14\rangle) \otimes V_3 |\phi\rangle +
\frac{1}{\sqrt{2}}(|03\rangle-|14\rangle) \otimes V_4 |\phi\rangle \\
&&+\frac{1}{\sqrt{2}}(|04\rangle+|12\rangle) \otimes V_5 |\phi\rangle +
\frac{1}{\sqrt{2}}(|04\rangle-|12\rangle) \otimes V_6 |\phi\rangle
 ),
\end{eqnarray*}
where $U_{1,2}=|0\rangle \langle 0|\pm |1\rangle \langle 1| +|2\rangle \langle 2|
+|3\rangle \langle 3|+|4\rangle \langle 4|$,
$U_{3,4}=|1\rangle \langle 0|\pm |0\rangle \langle 1| +|2\rangle \langle 2|
 +|3\rangle \langle 3|+|4\rangle \langle 4|$,
$V_{1,2}=|0\rangle \langle 2|+ |2\rangle \langle 0| \pm|1\rangle
\langle 3| \pm|3\rangle \langle 1| +|4\rangle \langle 4|$,
$V_{3,4}=|0\rangle \langle 3|+ |3\rangle \langle 0| \pm|1\rangle
\langle 4| \pm|4\rangle \langle 1| +|2\rangle \langle 2|$,
$V_{5,6}=|0\rangle \langle 4|+ |4\rangle \langle 0| \pm|1\rangle
\langle 2| \pm|2\rangle \langle 1| +|3\rangle \langle 3|$.
Obviously with respect to the Alice's measurement results,
faithful teleportation can be realized by applying the corresponding
unitary transformations $U_i$ and $V_j$, $i=1,...,4$,
$j=1,...,6$, on the Bob's part.

For mixed states $\rho$ in ${\bb C}^m \otimes {\bb C}^n$
if all its eigenstates belong to the class Eq. (\ref{mn-state}) and
any superpositions of these eigenstates belong to the class too, then $\rho$
can be used for faithful teleportation. For instance,
$\rho=p_1|\psi_1\rangle \langle \psi_1|+p_2|\psi_2\rangle \langle
\psi_2|$ with $|\psi_1\rangle =\frac{1}{2}(|00\rangle
+|11\rangle)+\frac{1}{\sqrt{6}}(|22\rangle+|33\rangle+|44\rangle)$,
$|\psi_2\rangle =\frac{1}{2}(|00\rangle
+|11\rangle)+\frac{1}{2}(|52\rangle+|63\rangle)$, $p_1+p_2=1$, can
be used for perfect teleportation of one qubit state.

{\it Remark.}~
In \cite{C. Y. Cheung} multipartite entangled states are used in faithful
teleportation of $d$-qubit state.
If we treat such teleportation as the teleportation of pure states in ${\bb
C}^{2^d}$, then it is easy to verify that the shared resources belong to
the class of states Eq. (\ref{mn-state}).

In the following we discuss the relations between the degree of entanglement
of the resource state $\rho$ and faithful teleportation.
Here we use the well-known
entanglement measure, entanglement of formation \cite{wootters}
to characterize the entanglement.
For a pure bipartite state $|\psi\ra_{AB}$, the entanglement of
formation is defined as the von Neumann entropy of either of the two
subsystems $A$ and $B$: $E(|\psi\ra_{AB})=-tr(\rho_{A(B)}\log_2\rho_{A(B)})$,
where $\rho_{A(B)}=tr_{B(A)}(|\psi\ra_{AB}\la\psi|)$ are the reduced density matrices.
For mixed state $\rho$ with pure state decompositions
$\label{decomposition}\rho=\sum_ip_i|\phi_i\ra\la\phi_i|$, $\sum_ip_i=1$, the
entanglement of formation is defined as the average entanglement of
the pure states in the decomposition, minimized over all possible pure state
decompositions of $\rho$: $E(\rho)=\textrm{inf}\sum_ip_iE(|\phi_i\ra)$.

It can be verified that either the maximally entangled pure state
$|\psi\ra$ in ${\bb C}^d\otimes {\bb C}^d$, or the mixed maximally
entangled state $\rho$ in ${\bb C}^{m}\otimes {\bb C}^d$, the
entanglement of formation is $E=\log d$. And entangled
states with entanglement of formation $E=\log d$ in these vector
spaces must be pure or mixed maximally entangled
\cite{lizhao}. Hence one has that the states in ${\bb C}^m\otimes
{\bb C}^d$ ($m\geq d$) can be used for faithful teleportation if and
only if their entanglement of formation is $\log d$. If Bob is
allowed to apply the measurements before the traditional teleportation protocol,
the states in ${\bb C}^d\otimes {\bb C}^n$ ($n\geq d$) can be
used for faithful teleportation if and only if their entanglement of
formation is $\log d$. For states $\rho$ in ${\bb
C}^m\otimes {\bb C}^n$ with $m,n>d$, the entanglement of formation
of $\rho$ presented in this paper for faithful teleportation is
larger than or equal to $\log d$. One may conjecture that
the entanglement of formation for all ideal
entangled resources is not less than $\log d$, giving rise to
another necessary condition for quantum states to be used for faithful
teleportation.

\section {Conlusions}

We have investigated systematically the necessary conditions
of entangled resource state ${\bb C}^m \otimes
{\bb C}^n$ $(m,n \geq d)$ for faithful teleportation of
$|\phi\ra$ in ${\bb C}^d$. It has been shown that for $n=d$, $\rho$ can
be used for faithful teleportation if and only if it is maximally
entangled. For $m=d$, $\rho$ can be used for faithful teleportation
if and only if it is a maximally entangled pure
state. For $m,n>d$, we present a class of pure and mixed states that
can be used for faithful teleportation. From the
point of view of experimental implementation of quantum
teleportation \cite{tel}, our results may help to understand the
character of faithful teleportation and to facilitate the experimental
implementations.

\noindent{\bf Acknowledgments}\, M. J. Zhao thanks
Max-Planck-Institute for Mathematics in the Sciences, Leipzig for
the hospitality. This work is supported by the NSFC 10875081, NSFC
10871227, KZ200810028013, PHR201007107 and NSF of Beijing 1092008.

\end{document}